\documentclass[aps,prc,preprint,amsmath,amssymb,showpacs,preprintnumbers,superscriptaddress]{revtex4-1}
\usepackage{graphicx}% Include figure files
\usepackage{dcolumn}% Align table columns on decimal point
\usepackage{bm}% bold math
\usepackage{color}

\usepackage{hyperref}% add hypertext capabilities

\begin{document}
%\begin{CJK*}{UTF8}{} % Use default fonts from CJK (see below)

\title{Interplay between antimagnetic and collective rotation in $^{58}$Fe}

\author{J. Peng  %({\CJKfamily{gbsn} 彭婧 }) 
}
         \affiliation{Department of Physics, Beijing Normal University, Beijing 100875, China}

\author{P. W. Zhao %({\CJKfamily{gbsn} 赵鹏巍 })
}
        \email{pwzhao@pku.edu.cn}
        \affiliation{Physics Division, Argonne National Laboratory, Argonne, Illinois 60439, USA}
        \affiliation{State Key Lab of Nuclear Physics {\rm\&} Technology, School of Physics, Peking University, Beijing 100871, China}

 \author{S. Q. Zhang %({\CJKfamily{gbsn} 张双全 })
}
         \affiliation{State Key Lab of Nuclear Physics {\rm\&} Technology, School of Physics, Peking University, Beijing 100871, China}

\author{J. Meng %({\CJKfamily{gbsn} 孟杰 })
}
         \affiliation{State Key Lab of Nuclear Physics {\rm\&} Technology, School of Physics, Peking University, Beijing 100871, China}
         \affiliation{School of Physics and Nuclear Energy Engineering, Beihang University, Beijing 100191, China}
         \affiliation{Department of Physics, University of Stellenbosch, Stellenbosch, South Africa}

\begin{abstract}
The self-consistent tilted axis cranking covariant density functional theory based on the point-coupling interaction PC-PK1
is applied to investigate the possible existence of antimagnetic rotation in the nucleus $^{58}$Fe.
The observed data for Band 3 and Band 4 are reproduced well with two assigned configurations.
It is found that both bands correspond to a rotation of antimagnetic character, but, due to the presence of considerable deformation,
the interplay between antimagnetic rotation and collective motion plays an essential role. In particular for Band 4, collective rotation is dominant in the competition with antimagnetic rotation. Moreover, it is shown that the behavior of
the ratios between the dynamic moments of inertia and the $B(E2)$ values reflects
the interplay between antimagnetic and collective rotation.
\end{abstract}

\pacs{21.60.Jz, 21.10.Re, 23.20.-g, 27.40.+z}
% 21.60.Jz Nuclear Density Functional Theory and extensions
%21.10.-k Properties of nuclei; nuclear energy levels
%21.10.Re Collective levels
%21.60.Ev Collective models
%21.60.Cs Shell model
%23.20.-g Electromagnetic transitions
%23.20.Js Multipole matrix element
%27.20.+n  6 A 19
%27.60.+j 90  A 149
%27.50.+e 59  A  89
%27.40.+z 39 A  58
%\keywords{}

\maketitle

%\end{CJK*}


%=======================================================================================
\section{Introduction}

The most common collective excitation in nuclei corresponds to a rotation about the principal axis of
the density distribution with the largest moment of inertia.
The substantial deformation of the overall density distribution specifies the orientation of the nucleus and, thus,
the rotational degree of freedom. In this picture, nuclear rotation is collective
and results from a coherent motion of many nucleons~\cite{Bohr1975}.

However, since the nucleons which form a nucleus carry a quantized amount of angular momentum,
the interplay between single-nucleon and collective motions is important in describing actual rotational excitations~\cite{Bohr1975}.
Moreover, such interplay leads to a variety of new phenomena according to the discrete symmetries obtained by combining the overall-deformation and the single-nucleon angular momentum vectors~\cite{Frauendorf2001Rev.Mod.Phys.463}.
For instance, in axially deformed nuclei, the coupling between the collective angular momentum and several valence holes
usually results in the high-$K$ bands as the collective angular momentum increases, and this gives rise to
well-known $K$ isomerism~\cite{Walker1999Nature35}.
In addition, a triaxial nucleus allows more degrees of freedom for the coupling between collective
and single-nucleon motions and, thus, is responsible for many new interesting modes such as nuclear chirality~\cite{Frauendorf1997Nucl.Phys.A131}, or longitudinal and transverse wobbling~\cite{Frauendorf2014Phys.Rev.C14322,Chen2014Phys.Rev.C44306,Matta2015Phys.Rev.Lett.82501}.

The situation for nearly spherical nuclei, however, is quite novel, since here collective rotation could be very weak due to the small deformation, while the valence nucleons play a crucial role in the generation of angular momentum.
The so-called magnetic and antimagnetic rotations are two typical examples as they are attributed to the gradual alignment of two angular momentum vectors of valence particles and/or holes with a specific orientation~\cite{Frauendorf2001Rev.Mod.Phys.463}.
For magnetic rotation, the energy and the angular momentum increase in terms of the shears mechanism; i.e., the alignment of the high-$j$ proton and neutron angular momenta~\cite{Frauendorf1993Nucl.Phys.A259}. The antimagnetic rotation, however, corresponds to the so-called ``two-shears-like mechanism''; i.e., the two blades of protons or neutrons are aligned back to back at the bandhead, and then simultaneously close with respect to each other while generating the total angular momentum~\cite{Frauendorf2001Rev.Mod.Phys.463}. Moreover, it has been demonstrated that the two novel magnetic and antimagnetic rotation modes can coexist in the same nucleus~\cite{Peng2015Phys.Rev.C44329}.

Since magnetic and antimagnetic rotations were proposed, lots of effort have been made to
understand these new phenomena and explore this manifestation throughout the nuclear
chart~\cite{Clark2000Annu.Rev.Nucl.Part.Sci.1,Frauendorf2001Rev.Mod.Phys.463,Hubel2005Prog.Part.Nucl.Phys.1}.
Up to now, more than 200 magnetic rotational bands spread in the mass regions of $A \sim 60$, $A \sim 80$, $A \sim 110$, $A \sim 140$,
and $A \sim 190$~\cite{Meng2013FrontiersofPhysics55} have been identified, while the antimagnetic rotational bands have been observed mainly
in Cd isotopes including $^{105}$Cd~\cite{Choudhury2010Phys.Rev.C61308}, $^{106}$Cd~\cite{Simons2003Phys.Rev.Lett.162501}, $^{107}$Cd~\cite{Choudhury2013Phys.Rev.C034304}, $^{108}$Cd~\cite{Simons2005Phys.Rev.C24318,Datta2005Phys.Rev.C41305}, $^{109}$Cd~\cite{Chiara2000Phys.Rev.C34318}, $^{110}$Cd~\cite{Roy2011Phys.Lett.B322} and, very recently,
in $^{101}$Pd~\cite{Sugawara2012Phys.Rev.C34326,Sugawara2015Phys.Rev.C24309}, $^{104}$Pd~\cite{Rather2014Phys.Rev.C061303(R)} and $^{143}$Eu~\cite{Rajbanshi2015.arXiv:nucl-th/1505.06074v1}.
Despite this evidence, it is found that pure magnetic and antimagnetic rotations are hardly realized in real nuclei.
In most of the observations, the contribution from a collective rotation mode is not negligible, and it always impacts the magnetic or antimagnetic rotation modes. In particular, for an antimagnetic rotation,
the rotational states are connected by $E2$ transitions only and the rotational axis is always along
one of the principal axes, and such features are also expected for a collective rotation.
Therefore, the investigation of the interplay between antimagnetic and collective rotations is important for a suitable description of observed antimagnetic rotational bands.

One of the most widely used models is a simple phenomenological one~\cite{Clark2000Annu.Rev.Nucl.Part.Sci.1}, where the competition between the two-shears-like mechanism and core rotation has been investigated, based on simple angular momentum geometry to be fitted to the data.
It is evident that a full understanding requires self-consistent microscopic investigations including all relevant degrees of freedom while based on reliable theories without additional parameters.
Such calculations are feasible in the framework of cranking density functional theories (DFTs).
In particular, the tilted axis cranking DFTs can explicitly construct the angular momentum vector diagrams showing the ``two-shears-like mechanism'', which is of great help
in visualizing the structure of antimagnetic rotational bands.
Recently, the tilted axis cranking covariant DFT~\cite{Zhao2011Phys.Lett.B181}
has successfully provided the first fully self-consistent and microscopic investigation of antimagnetic rotation~\cite{Zhao2011Phys.Rev.Lett.122501,Zhao2012Phys.Rev.C54310}. Note that the tilted axis cranking covariant DFT is not limited only to the description of antimagnetic rotation. It has been applied equally well to
magnetic rotation~\cite{Zhao2011Phys.Lett.B181,Yu2012Phys.Rev.C24318}, high-$K$ bands~\cite{zhao2015}, rotations with an exotic rod shape~\cite{Zhao2015Phys.Rev.Lett.22501}, etc.

Despite recent experimental and theoretical efforts, the study of antimagnetic rotation in the $A \sim 60$ mass region is still sparse.
Using heavy-ion induced fusion-evaporation reaction at Gammasphere~\cite{Steppenbeck2012Phys.Rev.C44316},
the newly observed high-spin states in the nucleus $^{58}$Fe provide an opportunity to investigate antimagnetic rotation in a light system. Note that in the previous study of Ref.~\cite{Steppenbeck2012Phys.Rev.C44316}, the presence of a magnetic rotational band had been suggested based on the tilted axis cranking covariant density functional theory (TAC-CDFT) with the configuration $\pi f^{-2}_{7/2} \otimes \nu[g^{1}_{9/2}(fp)^3]$~\cite{Steppenbeck2012Phys.Rev.C44316},
where two $f_{7/2}$ proton holes are aligned.
If these two $f_{7/2}$ proton holes are paired, it is easy to form the high-$j$ configuration and the angular momentum arrangement for antimagnetic rotation.
In this work, we will search for the possible existence of antimagnetic rotation in $^{58}$Fe.
We are focusing on Bands 3 and 4 of Ref.~\cite{Steppenbeck2012Phys.Rev.C44316},  which are $\Delta{I} = 2$
sequences. Band 3 has negative parity, has been seen in the $I\sim6$-$16\hbar$ range above a 5 MeV excitation energy and Band 4 has positive-parity, a $I\sim10$-$14\hbar$ range above 8 MeV excitation energy.
The level schemes, the relation between the rotational frequency
and the angular momentum as well as the dynamic ${\cal{J}}^{(2)}$ moment of inertia are calculated and compared with the available data.
Moreover, the interplay between the antimagnetic and collective modes will be discussed in details through the two-shears-like mechanism, the electromagnetic transition strengths $B(E2)$ and the ${\cal{J}}^{(2)}/B(E2)$ ratios.

%=======================================================================================
\section{Theoretical Framework}

In the TAC-CDFT, a rotating nucleon state is described by the Dirac equation in the rotating frame
\begin{equation}
\left[  \mbox{\boldmath$\alpha$}\cdot(-i\mbox{\boldmath$\nabla$}-\mbox{\boldmath$V$}%
)+\beta\left(  m+S\right)  +V-\mbox{\boldmath$\omega$}\cdot\hat{\mbox{\boldmath$J$}%
}\right]  \psi_{i}\;=\;\varepsilon_{i}\psi_{i}, \label{trmf-sneq}%
\end{equation}
where $\hat{\mbox{\boldmath$J$}}=\hat{\mbox{\boldmath$L$}}+\frac{1}{2}\hat{\mbox{\boldmath$\Sigma$}}$ is the total angular momentum of the nucleon spinors, and $S(\mathbf{r})$ and $V^{\mu}(\mathbf{r})$ are the relativistic scalar and vector fields, respectively, which are in turn coupled with the densities and currents. For more details, see Refs.~\cite{Zhao2011Phys.Lett.B181,Zhao2012Phys.Rev.C54310}.

In the present work, the Dirac equation is solved in a set of three dimensional harmonic oscillator bases with 10
major shells. The point-coupling density functional PC-PK1~\cite{Zhao2010Phys.Rev.C54319} is adopted, while the pairing correlations are neglected. The self-consistent rotational angle is determined by requiring that
$\mbox{\boldmath$\omega$}$ is parallel with $\mathbf{J}$ at fixed $\omega$. For antimagnetic rotation, this automatically leads to zero rotational angle; i.e., the rotational axis points always along the $x$ axis.

For Fe isotopes, the proton holes in the $f_{7/2}$ orbital and the neutron particles in the $g_{9/2}$ orbital
could form the matched high-$j$ configurations for antimagnetic rotation. In the present TAC-CDFT calculation, we adopt the valence nucleon configuration
$\pi f^{-2}_{7/2} \otimes \nu[g^{1}_{9/2}(fp)^3]$ for Band 3, and the configuration $\pi f^{-2}_{7/2} \otimes \nu[g^{2}_{9/2}(fp)^2]$ for Band 4,
where two $f_{7/2}$ proton holes are paired.
This configuration assignment is consistent with that proposed in the previous work~\cite{Steppenbeck2012Phys.Rev.C44316}, but in the latter the two $f_{7/2}$ proton holes for Band 4 are not fully paired.
For simplicity, the notations of Config 1 and Config 2 will be used to denote these two configurations hereafter.
In Table~\ref{tab:1}, the valence nucleon and the corresponding unpaired nucleon
configurations as well as their deformation parameters are listed for the ground state, Config 1 and Config 2.
The $\beta$ and $\gamma$ values for Config 1 and Config 2 shown here are only the values at the bandhead, i.e., at $\hbar\omega=0.2$ MeV. One can see that the $\beta$ deformations are not small for both Config 1 and Config 2, and this suggests that collective rotation might play an important role.

\begin{table}[h!tbp]
\centering\setlength{\tabcolsep}{0.35em}\fontsize{10pt}{11pt}\selectfont
 \caption{ Valence nucleon and unpaired nucleon configurations as well as the corresponding deformation parameters $\beta$ and $\gamma$. }
 \begin{tabular}{  c c c c c}
 \hline \hline
 \multicolumn{1}{c }{{\hspace{0.1cm}} notation {\hspace{0.1cm}} }
  &\multicolumn{1}{c@{\hspace{0.1cm}} }{{\hspace{0.1cm}} Valence~nucleon~configuration{\hspace{0.1cm}} }
  & Unpaired~nucleon~configuration    & $\beta$     & $\gamma$ \\ \hline\hline
      ground state        & $\pi f^{-2}_{7/2} \otimes \nu(fp)^4$                       &-                               &0.24&15.8$^\circ$\\ \cline{1-5}
      Config 1     & $\pi f^{-2}_{7/2} \otimes \nu[g^{1}_{9/2}(fp)^3]$          &$\nu[g^{1}_{9/2}(fp)^3]$ &0.28&14.3$^\circ$\\ \cline{1-5}
      Config 2     & $\pi f^{-2}_{7/2} \otimes \nu[g^{2}_{9/2}(fp)^2]$          &$\nu[g^{2}_{9/2}(fp)^2]$ &0.34&~1.4$^\circ$\\
  \hline\hline
 \end{tabular}
 \label{tab:1}
\end{table}

\section{Results and Discussion}

\begin{figure*}[!htbp]
\includegraphics[width=15cm]{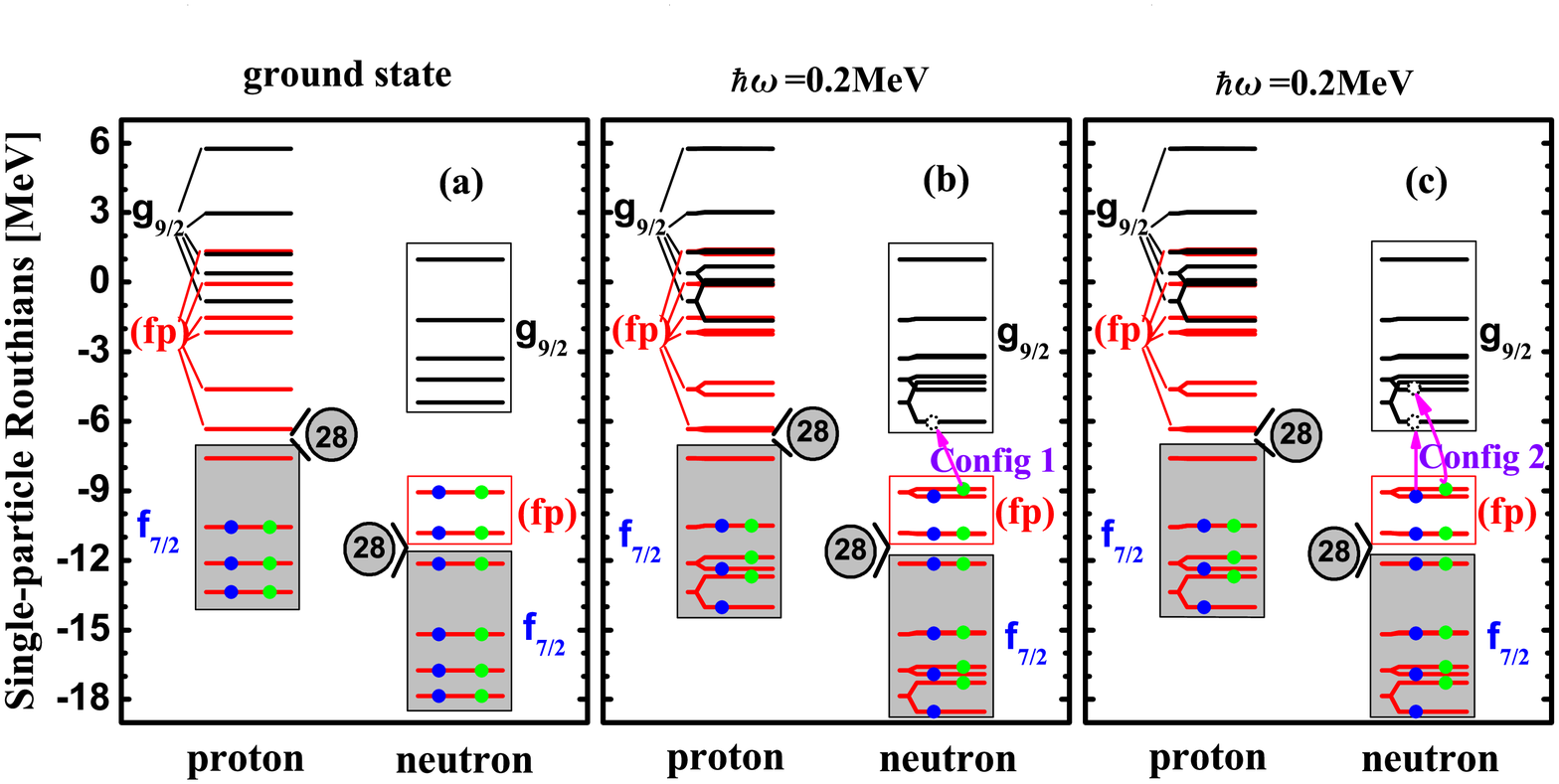}
\caption{(Color online) Schematic occupations for the configurations of the ground state (a), Config 1 (b) and Config 2 (c). Blue and green dots denote the occupied orbitals which are time-reversal conjugate states .
The single-particle Routhians with the ground-state configuration at $\hbar\omega=0.0$ MeV [panel (a)] and $\hbar\omega=0.2$ MeV [panels (b) and (c)] are presented for convenience. }
\label{Figocc}
 \end{figure*}

In the present TAC-CDFT calculation, as shown in Fig.~\ref{Figocc}(a), we first solve the Dirac equation for the ground state with $\hbar\omega=0.0$ MeV, by filling at each step of the iteration the proton and neutron
levels according to their energy from the bottom of the well.
As shown in Fig.~\ref{Figocc}(a), for the ground state, there are two paired proton holes sitting at the top of the $f_{7/2}$ shell, and four paired $(fp)$ neutrons above the $N = 28$ shell, and this configuration is associated with a triaxial deformation with $\beta=0.24$ and $\gamma=15.8^\circ$.
Then, we start to rotate the nucleus, and thus the time-reversal symmetry is violated by the Coriolis term in the Dirac equation, which leads to an energy splitting of the time-reversal conjugate states with an amplitude up to 1.75 MeV at $\hbar\omega=0.2$ MeV.
Note that a large splitting usually happens for states with a large expectation value $|j_x |$ which would induce strong Coriolis effects.
At $\hbar\omega=0.2$ MeV, the ground-state configuration; i.e., $\pi f^{-2}_{7/2} \otimes \nu(fp)^4$, is still shown as being the yrast one.
However, with increasing $\hbar\omega$, the large energy splitting between the time-reversal conjugate states allows a one-particle-one-hole neutron excitation from the $(fp)$ shell to the lowest $g_{9/2}$ orbital ($m_x=+\frac{9}{2}$), which leads to the configuration of $\pi f^{-2}_{7/2} \otimes \nu[g^{1}_{9/2}(fp)^3$] (Config 1, see Fig.~\ref{Figocc}(b)).
Similarly, the configuration $\pi f^{-2}_{7/2} \otimes \nu[g^{2}_{9/2}(fp)^2]$ is connected with a two-particle-two-hole neutron excitation from the $(fp)$ shell to the two lowest $g_{9/2}$ orbitals ($\nu g_{9/2},m_x=+\frac{9}{2}; \nu g_{9/2},m_x=+\frac{7}{2}$) (Config 2, see Fig.~\ref{Figocc}(c)).
In the following calculations with Config 1 or Config 2, the occupation of the valence nucleons is traced
at different rotational frequencies by adopting the same prescription as in Ref.~\cite{Peng2008Phys.Rev.C24313}.

\begin{figure}[!htbp]
 \includegraphics[width=8cm]{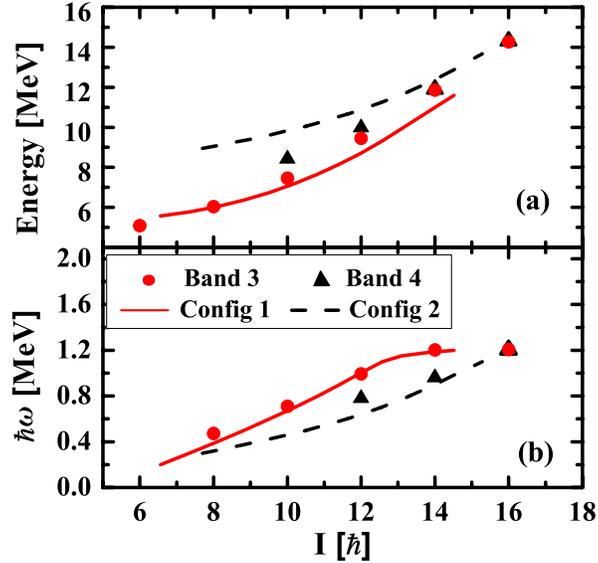}
 \caption{(Color online) Calculated energy spectra [panel (a)] and rotational frequency [panel (b)] with the configurations Config 1 (solid line) and Config 2 (dashed line) in comparison with the data for Band 3 (filled circles) and Band 4 (filled triangles)~\cite{Steppenbeck2012Phys.Rev.C44316} in $^{58}$Fe. The energy at $I = 8\hbar$ is taken as the reference in panel (a).} \label{Figiw}
\end{figure}

The calculated energy spectra and rotational frequency as functions of the total angular momentum are given in Fig.~\ref{Figiw} in comparison with the data for the observed Bands 3 and 4~\cite{Steppenbeck2012Phys.Rev.C44316} in $^{58}$Fe. In Fig.~\ref{Figiw}(a), one can clearly see that the excitation energies for Band 3 are reproduced well by the present TAC-CDFT calculation with Config 1. Moreover, the energies for the higher spin part of Band 4 are in good agreement of the calculated results for Config 2. Converged results were obtained up to around $I = 14.5\hbar$ for Config 1, and around 15$\hbar$ for Config 2.

Fig.~\ref{Figiw}(b) indicates clearly that the angular momenta for Band 3 and Band 4 are reproduced well by the calculations with Config 1 and Config 2, respectively, and this indicates also that the present calculation is able to reproduce the moments of inertia rather well.
For Band 3, in particular, the angular momentum increases almost linearly with the rotational frequency up to $I = 14\hbar$, while the observed unbending at $I=14\hbar$ may result from a level crossing between the proton $f_{7/2}$ and $(fp)$ orbitals.
In addition, it should be noted that the assigned configurations, Config 1 and Config 2,
have negative and positive parity, respectively, and this is consistent with the previous investigation,
based on the projected shell model~\cite{Steppenbeck2012Phys.Rev.C44316}.

\begin{figure}[!htbp]
\includegraphics[width=8cm]{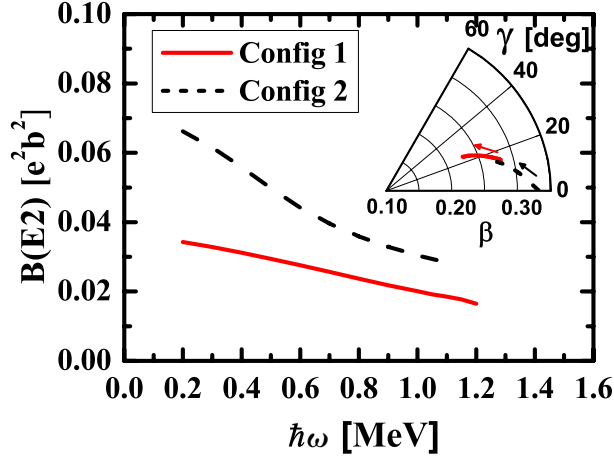}
\caption{(Color online) $B(E2)$ values as functions of the rotational frequency for Config 1 and Config 2.
Inset: Evolution of the deformation parameters $\beta$ and $\gamma$ driven by rotation with increase in frequency indicated by arrows.}
\label{Figbe2}
\end{figure}

Typical characteristics of antimagnetic rotation include the absence of $M1$ transitions and the decrease of the weak $E2$ transitions with spin. In Fig.~\ref{Figbe2}, the calculated reduced transition probabilities $B(E2)$, for Config 1 and Config 2
are presented as functions of the rotational frequency.
It is found that the calculated $B(E2)$ values here are very small ($<0.1~e^2b^2$), and they exhibit a smooth decrease with the growth in spin.
There are, at present, no available experimental $B(E2)$ values for these high-spin states in $^{58}$Fe.
Further measurements are welcome to validate the predicted electromagnetic transition probabilities, and this would be very useful to understand the nature of these two bands.
Furthermore, we note that the calculated $B(M1)$ values for both Config 1 and Config 2 vanish, and this is attributed to the fact that the transverse magnetic moments of two valence protons are anti-aligned and cancel out.

The decreasing tendencies of the $B(E2)$ values are connected with deformation changes.
As shown in the inset of Fig.~\ref{Figbe2}, with increasing rotational frequency, the nucleus undergoes a rapid decrease of $\beta$ deformation from 0.28 to 0.22 for Config 1, and from 0.34 to 0.26 for Config 2. Meanwhile, the $\gamma$ values keep increasing from $15^\circ$ to $23^\circ$ for Config 1, and from nearly $0^\circ$ to $16^\circ$ for Config 2.
Since considerable deformations are obtained here, it should be expected that collectivity plays a crucial role in the high-spin structure of these bands. Moreover, considering the fact that nuclear deformation here is changing toward stronger triaxiality with spin, this is even reminiscent of a ``band terminating'' picture that would apply to a normal collective band~\cite{Wadsworth1998Phys.Rev.Lett.1174}.

\begin{figure}[!htbp]
  \includegraphics[width=8cm]{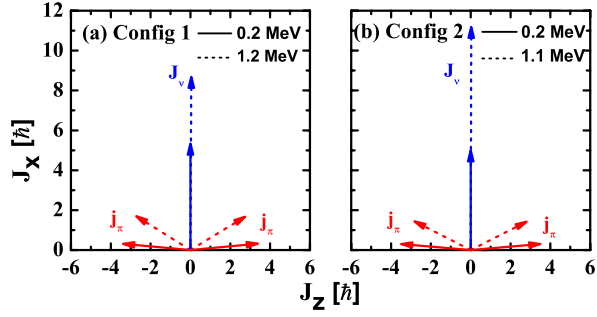}
  \caption{(Color online) Angular momentum vectors for all neutrons ${\mbox{\boldmath$J$}}_{\nu}$ (blue), and the two proton holes in the $f_{7/2}$ shell ${\mbox{\boldmath$j$}}_\pi$ (red) for Config 1 (a) and Config 2 (b). }
  \label{Figshears}
\end{figure}

Therefore, it becomes important to check the mechanism behind the generation of the angular momentum in Bands 3 and 4.
We present in Fig.~\ref{Figshears} the angular momentum vectors for all neutrons ${\mbox{\boldmath$J$}}_{\nu}$ and the two proton holes, in the $f_{7/2}$ shell, ${\mbox{\boldmath$j$}}_\pi$ for Config 1 and Config 2, respectively.
Here, the symbols have the same definition as in Refs.~\cite{Zhao2011Phys.Rev.Lett.122501,Zhao2012Phys.Rev.C54310}.
At the bandhead ($\hbar\omega = 0.2$ MeV), the two proton angular momentum vectors ${\mbox{\boldmath$j$}}_\pi$ are pointing opposite to each other, and they are nearly perpendicular to the neutron angular momentum vector ${\mbox{\boldmath$J$}}_{\nu}$ for both configurations.
Together with ${\mbox{\boldmath$J$}}_{\nu}$, they form the blades of the two shears.
With increasing rotational frequency, the gradual alignment of the vectors ${\mbox{\boldmath$j$}}_\pi$ toward the vector ${\mbox{\boldmath$J$}}_{\nu}$ generates partially the total angular momentum, and this corresponds to the so-called ``two-shears-like mechanism'', where the two shears are closing simultaneously.
On the other hand, it should be noted that the increase of the neutron angular momentum also contributes notably to the generation of the total angular momentum along the bands.
In particular for the Config 2 band, the neutron angular momentum jumps by more than $6\hbar$ when the rotational frequency increases from 0.2 MeV to 1.1 MeV, while the alignment of the two proton holes only provides a contribution of less than $4\hbar$.
The increase of the neutron angular momentum results from the contribution of collective rotation.
Thus, this is an instance where one can clearly see the interplay between the antimagnetic and collective rotation in the two configurations. 

\begin{figure}[!htbp]
\includegraphics[width=8cm]{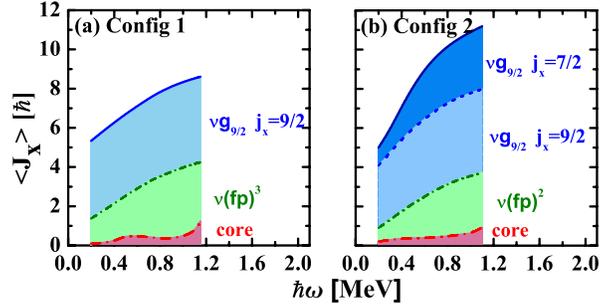}
\caption{(Color online) Contributions of the neutrons in the $g_{9/2}$ orbital, in the $(fp)$ shell, and the neutron $N=28$ ``core" to the total neutron angular momentum along the $x$ axis for Config 1 (a) and Config 2 (b). }
\label{Figjx}
\end{figure}

In a microscopic picture, the angular momentum comes from the individual nucleons self-consistently.
Therefore, it is of interest to extract the contributions of the individual neutrons to
the neutron angular momentum, and these are found in Fig.~\ref{Figjx} for Config 1 and Config 2, respectively.
One can see that the angular momentum contribution from the neutron $N=28$ core; i.e.,
from all the orbitals below $N=28$, is quite small ($\lesssim 1\hbar$) for Config 1 and Config 2.
However, the neutron angular momenta are almost all coming from the four valence neutrons in the $g_{9/2}$ and $(fp)$ orbitals, which corresponds to a ``band termination''.

For Config 1, a neutron sitting at the bottom of the $g_{9/2}$ shell contributes an angular momentum of roughly $4\hbar$.
When the rotational frequency increases from 0.3 MeV to 1.2 MeV, the contribution of this neutron does barely change, and the increment of the angular momentum comes mostly from the alignment of the other three neutrons in the $(fp)$ shell, which is driven by the collectivity.
In regards to the generation of the total angular momentum along the $x$ axis, 
the alignment of the neutron angular momentum contributes 3.35$\hbar$ (55\%), while the two-shears-like mechanism contributes 2.74$\hbar$ (45\%).

For Config 2, one additional valence neutron occupies the $g_{9/2}$ orbital. However, this neutron contributes almost nothing to the angular momentum at the bandhead, and this indicates that there is very strong mixing between this orbital and the low-$j$ $(fp)$ orbits.
Along the band, this $g_{9/2}$ orbital becomes purer and purer, and eventually provides roughly 3$\hbar$ to the total angular momentum. Therefore, such alignment originates from collective rotation as well, and together with the alignment of around 2.81$\hbar$, obtained from the two neutrons in the ($fp$) shell, collective rotation then accounts for almost 73\% to the increment of the total angular momentum.

Therefore, one can conclude that both the Config 1 and Config 2 bands correspond to rotating bands of the antimagnetic rotation character, but due to the substantial deformation, the interplay between  antimagnetic rotation and collective motion plays an essential role.
In particular for the Config 2 band, collective rotation is dominant in the competition with antimagnetic rotation.
The higher angular momenta originate from the two-shears-like mechanism
of antimagnetic rotation together with the alignment generated by the collective rotation.

\begin{figure}[!htbp]
\includegraphics[width=8cm]{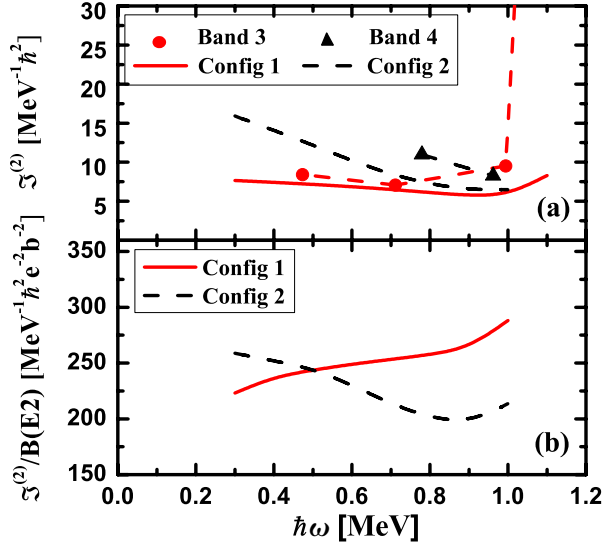}
\caption{(Color online) Dynamic moments of inertia ${\cal{J}}^{(2)}$ [panel (a)] and ${\cal{J}}^{(2)}/B(E2)$ ratios [panel (b)] as functions of the rotational frequency for Config 1 and Config 2 in comparison with the data (circles and triangles), when available~\cite{Steppenbeck2012Phys.Rev.C44316}.  }
\label{FigMOI}
\end{figure}

The complex interplay between the antimagnetic and collective rotations can be revealed by the dynamic moment of inertia ${\cal{J}}^{(2)}$ and by the ${\cal{J}}^{(2)}/B(E2)$ ratio.
These are given in Fig.~\ref{FigMOI} as functions of the rotational frequency, in comparison with the data available~\cite{Steppenbeck2012Phys.Rev.C44316}.
The abrupt rise in the experimental dynamic moment of inertia of Band 3 ($\hbar\omega \sim 1.0$ MeV) corresponds to the unbending, and has been discussed in Fig.~\ref{Figiw}.

It is found that the moments of inertia for both Config 1 and Config 2 are reproduced reasonably, especially their evolution with frequency.
For Config 1, the rise of the ${\cal{J}}^{(2)}/B(E2)$ ratios is characteristic of antimagnetic rotation, reflecting the fact that ${\cal{J}}^{(2)}$ is essentially constant whereas the $B(E2)$ values rapidly approach zero as the spin increases along the band (see Fig.~\ref{Figbe2}).
This is also consistent with the prediction of the classical model of Ref.~\cite{Simons2005Phys.Rev.C24318}, where the behavior of $B(E2)$ and ${\cal{J}}^{(2)}$ completely reflects how the interaction between the high-$j$ orbitals depends on their relative orientation. Note that for a pure antimagnetic rotor, the angular momentum is carried by only a few nucleons in high-$j$ orbitals.

For Config 2, however, the ${\cal{J}}^{(2)}/B(E2)$ ratios exhibit decreasing tendency along the band, indicating that both ${\cal{J}}^{(2)}$ and $B(E2)$ values drop with the reduction of deformation as the spin increases, and the former drop even faster than the latter. The rapid drop of ${\cal{J}}^{(2)}$ ($dI/d\omega$) can be understood from the alignment of $g_{9/2}$
neutrons driven by the collective rotation, as seen in Fig.~\ref{Figjx}(b), the increment of which occurs gradually with increasing rotational frequency. It should be kept in mind that collective rotation is a motion carried by many nucleons, each of which contributes a small
fraction to the total angular momentum.
The averaging over the individual nucleon contributions results in both the ${\cal{J}}^{(2)}$ and $B(E2)$  values being related to the deformation of the nuclear density distribution.

\section{summary}

In summary, the self-consistent tilted axis cranking covariant density functional theory based on a point-coupling interaction has been applied to investigate the possible existence of the antimagnetic rotation in the nucleus $^{58}$Fe.
The energy spectra, the relation between the spin and the rotational frequency, the dynamic moment of inertia, the deformation parameters and the reduced $E2$ transition probabilities have been studied.
The energy spectra of Band 3 have been reproduced well with the configuration $\pi f^{-2}_{7/2} \otimes \nu[g^{1}_{9/2}(fp)^3]$.
The observed energies for the higher spin part of Band 4 are in good agreement with the calculated results for the configuration $\pi f^{-2}_{7/2} \otimes \nu[g^{2}_{9/2}(fp)^2]$.
The absence of measurable $B(M1)$ strength and the decreasing $B(E2)$ values for both Config 1 and Config 2 are consistent with the picture of a two-shears-like mechanism, which has been demonstrated by the orientation of the two-proton-hole and neutron angular momenta.

However, due to the presence of considerable deformation, collective rotation provides a significant  contribution to the total angular momentum as well. In particular for the Config 2 band, collective rotation is dominant in the competition with antimagnetic rotation.
In both bands, the interplay between antimagnetic rotation and collective motion plays an essential role, and it can be revealed by the behavior of their ${\cal{J}}^{(2)}/B(E2)$ ratios with the spin.

\section*{Acknowledgements}

The authors are grateful to R. V. F. Janssens and Robert B. Wiringa for helpful discussions and critical reading of the manuscript. This work is supported by the Major State 973 Program of China (Grant No.
2013CB834400), the National Natural Science Foundation of China (Grants No. 11175002,
No. 11335002, No. 11461141002, No. 11105005), the Open Project Program of State Key Laboratory of Theoretical Physics, Institute of Theoretical Physics, Chinese Academy of Sciences, China (No.Y4KF041CJ1), and by U.S. Department of Energy (DOE), Office of Science, Office of Nuclear Physics, under contract DE-AC02-06CH11357.

%\bibliography{paper}


\end{document}